\documentclass[11pt]{iopart}
\usepackage{amssymb}
\usepackage[dvips]{graphics}
\newtheorem{thm}{Theorem}[section]
\newtheorem{prop}{Proposition}[section]

\newtheorem{lemma}{Lemma}[section]

\newtheorem{defn}[thm]{Definition}

\newtheorem{rem}[thm]{Remark}

\newcommand{\vege}{\hfill $\Box $}
\newcommand{\fimp}{\vspace{.1in}}
\newcommand{\dem}{\noindent{\bf Proof. }}
\newcommand{\real}{{\mathbb R}}
     
\begin{document}
\title[]{Symmetric flows and Darcy's law in Curved Spaces}
\author{Eduardo S. G. Leandro$^{1}$, Jos\'e A. Miranda $^{2}$\footnote{On leave from Departamento 
de F\'{\i}sica, UFPE, Brazil} and Fernando Moraes $^{3}$}
\address{
$^{1}${Departamento de Matem\'atica, Universidade Federal de Pernambuco, 
58670-901, Recife, PE, Brazil }.\\
$^{2}${Department of Physics, University of Florida, P. O. Box 118440, Gainesville, Florida 32611-8440}.\\
$^{3}${Departamento de F\'{\i}sica, Universidade Federal da
Para\'{\i}ba, Caixa Postal 5008, 58051-970, Jo\~ao Pessoa, PB, Brazil.}}
%\email{}\email{}

\begin{abstract}
 We consider the problem of existence of certain symmetrical solutions of Stokes equation on a three-dimensional manifold $M$ with a general metric possessing symmetry. These solutions correspond to unidirectional flows. We have been able to determine necessary and sufficient conditions for their existence. Symmetric unidirectional flows are fundamental for deducing the so-called Darcy's law, which is the law governing fluid flow in a Hele-Shaw cell embedded in the environment $M$. Our main interest is to depart from the usual, flat background environment, and consider the possibility of an environment of arbitrary constant curvature $K$ in which a cell is embedded. We generalize Darcy's law for particular models of such spaces obtained from $\real^3$ with a conformal metric. We employ the calculus of differential forms for a simpler and more elegant approach to the problems herein discussed.
\end{abstract}

\vspace{2pc}
%\submitto{\JPA}

\maketitle

\section{Introduction} \label{introduc}

Pattern formation is a very exciting and fastly growing area in physics and related sciences \cite{Rev1,Rev2,Rev3,Rev4,Rev5}. The Saffman-Taylor \cite{saff} problem is one  of the most studied among the systems presenting formation and evolution of patterned structures. It studies the hydrodynamic instabilities at the interface separating two immiscible fluids confined between two parallel flat plates, the Hele-Shaw cell. In such configuration, when a low-viscosity  fluid displaces a higher viscosity fluid, the interface becomes unstable, deforms, and forms fingers \cite{Ben,Hom,McC,flat}. The key point in the study of such patterns is Darcy's law, the two-dimensional reduction of the Navier-Stokes equation incorporating the boundary conditions (no-slip) and the mass conservation law (continuity equation). An interesting generalization of this problem is to vary the geometry of the cell and study how geometric parameters like curvature, for instance, affect the interface dynamics. Spherical \cite{sphere,sphereg}, cylindrical \cite{cyl} and conical \cite{cone} geometries have been studied in this context, yielding substantial information connecting relevant fingering mechanisms (finger competition and finger tip-splitting) to the cell's geometric and topological features. It is our aim here to provide the means of further generalizations of the Saffman-Taylor problem involving other geometries. 

In this work we derive Darcy's law in more general geometries, where we allow the ordinary three-dimensional space ${\mathbb R}^3$ to acquire a symmetrical Riemannian metric \cite{rimgeom}, then look for the conditions  wherein Navier-Stokes equation in this space becomes separable. An important case where this occurs is that of a separable metric. For this case, we demonstrate that symmetric unidirectional flows are possible. Another important case is that of ${\mathbb R}^3$ endowed with a conformal metric. For this case, we derive Darcy's law even in the situations where symmetric unidirectional flows do not exist. The case of a pseudo-Riemannian metric (Minkowski) is also approached for it supports surfaces of constant negative curvature (pseudospheres or Lobachevsky planes), which have been partially studied in \cite{sphere}. The calculus of differential forms \cite{darling} is used throughout the article for a simpler and more elegant way of presenting the problem.

Suppose $M$ is a smooth orientable manifold of dimension 3 endowed with a metric locally given by
\begin{equation} \label{metric}
ds^2= E_1^2 dx_1^2+ E_2^2dx_2^2+E_3^2 dx_3^2,
\end{equation}
where  $E_i$ are smooth functions of the coordinates $x_i$, $i=1,2,3$. Let $(\;,\;)$ be the inner product induced by this metric.

 The motion of a fluid in M is described by a vector field $\vec{V}: M \rightarrow TM$, where $TM$ denotes the tangent bundle to $M$. There exists a canonical correspondence between $TM$ and the cotangent bundle $TM^*=\bigwedge^1(TM^*)$, which is defined using the metric (\ref{metric}): to each tangent vector $\vec{V} \in TM$ there corresponds a unique differential 1-form $\omega_{\vec{V}}\in \bigwedge^1(TM^*) $ such that $\omega_{\vec{V}}(\vec{W})=(\vec{V},\vec{W})$. Given an orthonormal basis $\beta=\{\vec{e}_1,\vec{e}_2,\vec{e}_3\}$ of $TM$, we define the corresponding basis of $\bigwedge^1(TM^*)$ as $\beta^*=\{\omega_{\vec{e}_1},\omega_{\vec{e}_2},\omega_{\vec{e}_3}\}$. The basis $\beta^*$ is orthonormal in the inner product $\langle \;,\; \rangle$ induced by the dual metric
\begin{equation} \label{dualmetric}
  (ds^*)^2=(1/E_1^2) dx_1^2+(1/E_2^2) dx_2^2+(1/E_3^2) dx_3^2. 
\end{equation}

  Fluid motion is governed by the Navier-Stokes equation
\[ \rho \left[\frac{\partial \vec{V}}{\partial t}+(\vec{V},\nabla)\vec{V} \right]= -grad(p)+\eta \Delta \vec{V}. \]
If the flow is incompressible, $\vec{V}$ must also satisfy the equation of continuity
\[div \vec{V}=0.\]
In many applications, such as the study of Hele-Shaw flows, one assumes a steady flow and neglects the so-called inertial terms on the left-hand-side of the Navier-Stokes equation. Under these hypotheses, Navier-Stokes equation reduces to Stokes equation. Using the correspondence between vector fields and differential forms, Stokes equation  and the equation of continuity translate respectively into
\begin{eqnarray} \label{stokes}
-\omega_{grad(p)}+& \eta \Delta \omega_{\vec{V}}&  =  0,\\
& \delta \omega_{\vec{V}}& = 0,
\end{eqnarray}
where $\Delta$ is the Laplace operator 
\[-(d \delta+\delta d). \]
We recall that the operators $d$ and $\delta$ are the exterior differential and codifferential, respectively. The codifferential is an operator from $\bigwedge^k(TM^*)$ to $\bigwedge^{k-1}(TM^*)$ defined by
\[ \delta=(-1)^{k}*d*, \]
where $*:\bigwedge^k(TM^*) \rightarrow \bigwedge^{3-k}(TM^*)$ is the Hodge star operator.

 Darcy's law is obtained by averaging  $\vec{V}$ in the normal direction with respect to a given two-dimensional smooth submanifold $N$ of $M$. It provides a reasonable description of the fluid motion between two non-intersecting neighboring copies of $N$. Generally such pair of submanifolds is said to form a Hele-Shaw cell. Of particular interest, due to their simplicity, are the cells formed by level sets $\{x_i=constant\}$ in a local chart of $M$. All examples of Hele-Shaw cells studied so far (planar, cylindrical, conical, spherical) are formed by such submanifolds. One procedure for obtaining Darcy's law consists of considering a one-parameter family of velocity fields which corresponds to what we called symmetric unidirectional flows. When such family exists and its profile function (denoted in this paper by $g$) is non-constant, a simple method provides a quick deduction of Darcy's law. The existence of symmetric unidirectional flows is, therefore, a very important issue. One of our main goals is to attempt to overcome the non-existence of such flows in a perturbative way. 
 
 In Section 2 we look for solutions of Stokes equations for a symmetric unidirectional flow and analyze under which conditions the equation is separable. In Section 3 we find the conditions for the existence of symmetric flows in curved space and in Section 4 we study the solutions of Stokes equation (\ref{stokes}) and deduce Darcy's law for the following systems of the Hele-Shaw type:
\begin{itemize}
\item[(1)] Two nearby pseudo-spheres in Minkowski's 3-space;
\item[(2)] Two parallel planes in ${\mathbb R}^3$ with a conformal metric.
\end{itemize}
Section 5 summarizes our main results and conclusions.

\section{Solution of Stokes equation for a symmetric unidirectional flow}

 We study the solutions of Stokes equation in M under the following assumptions
\begin{itemize}
\item [(A1)] The level sets $S_a=\{x_1=a\}$ and $S_{a+b}=\{x_1=a+b \}$ are two smooth (non-intersecting) surfaces. The vector field $\vec{e}_1$ is normal to both $S_a$ and $S_{a+b}$.
\item [(A2)] The coefficients of the metric~(\ref{metric}) of $M$, do not depend on $x_2$.
\item [(A3)] Fluid motion is in the direction of $x_3$ and the velocity field does not depend on $x_2$. We refer to such motion as a \bf{\emph{symmetric unidirectional flow}}.
\end{itemize}

 A velocity field satisfying assumption (A3) has the form
\[ \vec{V}=V_{3}(x_1,x_3) \vec{e}_{3}. \]
Using the correspondence with 1-forms, we obtain \[ \omega_{\vec{V}}=V_3 \omega_{\vec{e}_3}=(V_3 E_3) d x_3.\]
The equation of continuity then writes
\[ \delta \omega_{\vec{V}}= -*d*(V_3 E_3 d x_3)=-*d(V_3 E_1 dx_1 \wedge E_2 d x_2)=-*(\frac{\partial(V_3 E_1 E_2)}{\partial x_3} dx_1 \wedge dx_2 \wedge dx_3)=0. \]
Thus we have $\frac{\partial(E_1 E_2 V_3)}{\partial x_3}=0$, which implies that
\begin{eqnarray} \label{g-eq}
 E_1 E_2 V_3=g(x_1)
\end{eqnarray}
for some function $g$ which remains to be determined. 

 Using the equation of continuity, the Laplacian of $\omega_{\vec{V}}$ reduces to $-\delta d \omega_{\vec{V}}$, i.e.
\begin{eqnarray*}
-*d*d(V_3 E_3 dx_3)&=&-*d*\left(\frac{\partial (V_3 E_3)}{\partial x_1} dx_1 \wedge  dx_3 \right)\\
&=&-*d\left(-\frac{E_2}{E_1E_3}\frac{\partial (V_3 E_3)}{\partial x_1}  dx_2 \right)
\end{eqnarray*}
which equals
\begin{eqnarray*}
&&*\left(\frac{\partial}{\partial x_1}\left(\frac{E_2}{E_1E_3}\frac{\partial (V_3 E_3)}{\partial x_1}\right) dx_1 \wedge dx_2-\frac{\partial}{\partial x_3}\left(\frac{E_2}{E_1E_3}\frac{\partial (V_3 E_3)}{\partial x_1}\right) dx_2 \wedge dx_3\right)\\
&=& -\frac{E_1}{E_2E_3}\frac{\partial}{\partial x_3}\left(\frac{E_2}{E_1E_3}\frac{\partial (V_3 E_3)}{\partial x_1}\right) dx_1+\frac{E_3}{E_1E_2}\frac{\partial}{\partial x_1}\left(\frac{E_2}{E_1E_3}\frac{\partial (V_3 E_3)}{\partial x_1}\right) dx_3.
\end{eqnarray*}
 
 Recall that, by definition of the gradient, 
\[\omega_{grad (p)}=dp=\frac{\partial p}{\partial x_1}dx_1+\frac{\partial p}{\partial x_3}dx_3.\]
Thus, from the expression for the Laplacian of  $\omega_{\vec{V}}$ deduced above, Stokes equation is equivalent to the following system of equations
\begin{eqnarray}
-\frac{\partial p}{\partial x_1}-\eta\frac{E_1}{E_2E_3}\frac{\partial}{\partial x_3}\left(\frac{E_2}{E_1E_3}\frac{\partial (V_3 E_3)}{\partial x_1}\right)&=&0, \label{stokes1}\\
-\frac{\partial p}{\partial x_3}+\eta\frac{E_3}{E_1E_2}\frac{\partial}{\partial \label{stokes2} x_1}\left(\frac{E_2}{E_1E_3}\frac{\partial (V_3 E_3)}{\partial x_1}\right)&=&0.
\end{eqnarray}
The unknown function $g(x_1)$ in (\ref{g-eq}) must be such that the system above is satisfied for some smooth function $p(x_1,x_3)$. If we apply the operator $d$ to both sides of Stokes equation
\[ -dp+\eta \Delta \omega_{\vec{V}}=0 \]
we obtain, since $\eta \neq 0$,
\[  d \Delta \omega_{\vec{V}}=0 \]
because $d^2=0$. Hence the (local) existence of $p$ satisfying Stokes equation is guaranteed if $g$ solves the equation
\[ d \Delta \omega_{\vec{V}}=-d\delta d\left(\frac{gE_3}{E_1E_2} \right)=0, \]
or, in coordinates, the equation
\begin{equation} \label{maineq}
 \frac{\partial}{\partial x_3}\left(\frac{E_1}{E_2E_3}\frac{\partial}{\partial x_3}\left(\frac{E_2}{E_1E_3}\frac{\partial \left(\frac{g E_3}{E_1 E_2}\right)}{\partial x_1}\right)\right)=\frac{\partial}{\partial x_1}\left(\frac{E_3}{E_1E_2}\frac{\partial}{\partial x_1}\left(\frac{E_2}{E_1E_3}\frac{\partial \left(\frac{g E_3}{E_1 E_2}\right)}{\partial x_1}\right)\right)
\end{equation}
which is a linear ordinary differential equation of third order whose coefficients are functions of the metric coefficients. Thus the existence of non-constant solutions of equation~(\ref{maineq}) depends on the metric. From now on we will only consider non-constant solutions of~(\ref{maineq}).

\subsection{ Separable Stokes Equation}

 A great simplification is achieved when the metric is such that Stokes equation is reduced to a single separable differential equation. It turns out that for some important examples Stokes equation reduces to a separable equation~(\ref{stokes2}). This fact motivates the definition below.

\begin{defn} \label{sepdef}
 We say that Stokes equation is {\bf\emph{separable}} if the system of differential equations~(\ref{stokes1}) and~(\ref{stokes2}) reduces to a separable equation~(\ref{stokes2}).
\end{defn} 
 
 The following definition will also be useful in our discussion.
 
\begin{defn}
 A function $f(x_1,x_3)$ is {\bf\emph{separable}} if it can be written as a product of a function of $x_1$ and a function of $x_3$.
\end{defn}

 Let us suppose that
\begin{equation} \label{sepsimp}
 \frac{\partial}{\partial x_3}\left(\frac{E_2}{E_1E_3}\frac{\partial }{\partial x_1}\left(\frac{g E_3}{E_1 E_2}\right)\right)=0,
\end{equation}
or, equivalently
\begin{equation} \label{sep}
 \frac{E_2}{E_1E_3}\frac{\partial }{\partial x_1}\left(\frac{g E_3}{E_1 E_2}\right)=C(x_1).
\end{equation}
for some function $C$.

 Multiplying both sides of the equation above by $g(x_1)$, and denoting $\frac{g}{E_1}$ by $G$, we obtain
\[G\,\left(\frac{E_2}{E_3}\right)\frac{\partial }{\partial x_1}\left(G\frac{E_3}{E_2}\right)=Cg . \]
Dividing both sides by $G^2$, we get
\begin{eqnarray*}
%G \frac{\partial G}{\partial x_1}+G^2 \frac{\partial }{\partial x_1}ln \left(\frac{E_3}{E_2}\right)=Cg 
% & \Leftrightarrow & \frac{1}{G}\frac{\partial G}{\partial x_1}+\frac{\partial }{\partial x_1}ln \left(\frac{E_3}{E_2}\right)=\frac{C g}{G^2} \\
 \frac{\partial }{\partial x_1}ln \left(\frac{G E_3}{E_2}\right)=\frac{C g}{G^2}
 & \Leftrightarrow &  ln \left(\frac{G E_3}{E_2}\right)=\int \frac{C g}{G^2}dx_1 +\tilde{C}(x_3), 
\end{eqnarray*}
which in terms of $g$ is
\begin{eqnarray*}
ln \left(\frac{g E_3}{E_1E_2}\right)= \int \frac{C E_1^2}{g}dx_1 +\tilde{C} & \Leftrightarrow &
\frac{E_3}{E_1E_2}=\frac{exp(\tilde{C} )exp \left(\int \frac{C E_1^2}{g}dx_1 \right)}{g}.
\end{eqnarray*}
The last equation above shows that equation~(\ref{sep}) holds as long as the ratio $\frac{E_3}{E_1E_2}$ is the product of a function of $x_1$, a function of $x_3$ and a function of $x_1$, $x_3$. 

\begin{prop} \label{stokessep}
Assume that $E_1=E_1(x_1)$. Then Stokes equation is separable if and only if the ratio $\frac{E_3}{E_2}$ is separable.
\end{prop}
\dem If $E_1=E_1(x_1)$ and if Stokes equation is separable in the sense of definition~\ref{sepdef}, then from equation~(\ref{sepsimp})
\[\frac{E_3}{E_1E_2}=\frac{exp(\tilde{C} )exp \left(\int \frac{C E_1^2}{g}dx_1 \right)}{g} \Rightarrow \frac{E_3}{E_2}=H(x_1)\tilde{H}(x_3) \]
for some functions $H,\tilde{H}$.
 
 On the other hand, if $\frac{E_3}{E_2}=H(x_1)\tilde{H}(x_3)$ then~(\ref{sepsimp}) holds and equation~(\ref{stokes2}) becomes
\[-\frac{\partial p}{\partial x_3}+\eta H(x_1)\tilde{H}(x_3)\frac{\partial}{\partial  x_1}\left(\frac{E_2}{E_1E_3}\frac{\partial (V_3 E_3)}{\partial x_1}\right)=0.\]
After dividing by $\tilde{H}(x_3)$ and moving the second term to the right hand side, we have 
\[ \frac{1}{\tilde{H}}\frac{\partial p}{\partial x_3}=\eta H\frac{\partial}{\partial  x_1}\left(\frac{E_2}{E_1E_3}\frac{\partial }{\partial x_1}\left(\frac{g E_3}{E_1 E_2}\right)\right)=\eta H\frac{\partial}{\partial  x_1}\left(\frac{1}{E_1H}\frac{\partial }{\partial x_1}\left(\frac{g H}{E_1 }\right)\right). \]
Clearly each side of the equation above depends on a single variable. \vege
\fimp

 We now discuss a few examples.\\

\noindent{\bf{\emph{Example 1:}}} Let $M={\mathbb R}^3$ with the Euclidean metric in cylindrical coordinates $(\rho, \phi, z)$. We have that
\begin{eqnarray*}
E_1=E_3=1, & E_2=\rho.
\end{eqnarray*}
From proposition~\ref{stokessep}, it follows that Stokes equation for a flow in the $z$-direction reduces to equation~(\ref{stokes2}) 
\begin{equation} \frac{\partial p}{\partial z}=\frac{\eta} {\rho} \frac{\partial}{\partial  \rho}\left(\rho\frac{\partial }{\partial \rho}\left(\frac{g}{\rho}\right)\right). 
\end{equation}
 
 If, instead, we consider flows in the $\rho$-direction (radial flow) which are $\phi$-symmetric, then Stokes equation is
\begin{equation}
\frac{1}{\rho} \frac{\partial p}{\partial \rho}=\eta g''. 
\end{equation}
Finally, for $z$-symmetric flows in the $\phi$-direction, we have

\begin{equation} \frac{\partial p}{\partial \phi}=\eta \rho \frac{\partial}{\partial  \rho}\left(\frac{1}{\rho}\frac{\partial }{\partial \rho}(\rho\, g)\right). 
\end{equation}
\fimp

\noindent{\bf{\emph{Example 2:}}} Let $M={\mathbb R}^3$ with the Euclidean metric in spherical coordinates $(r, \phi, \theta)$. In this case
\begin{eqnarray*}
E_1=1, &E_2=r \sin \theta ,& E_3=r.
\end{eqnarray*}
Proposition~\ref{stokessep} implies that Stokes equation for polar flows reduces to 
\begin{equation} 
 \sin \theta \frac{\partial p}{\partial \theta}=\eta \,g''. 
\end{equation}
\fimp
\noindent{\bf{\emph{Example 3:}}} Our arguments so far apply to semi-Riemannian metrics just as well. Let $M={\mathbb R}^3$ with the Minkowski metric 
\[ds^2_M=-dr^2+r^2 \sinh^2\tau d\phi^2+r^2 d\tau^2  \]
in pseudo-spherical coordinates  $(r, \phi, \tau)$.
The analogous of Proposition~\ref{stokessep} for a semi-Riemannian metric implies that Stokes equation reduces to 
\begin{equation} \label{mink}
 \sinh \tau \frac{\partial p}{\partial \tau}=\eta \,g''. 
\end{equation}

\subsection{An example of non-separable Stokes equation: flow in ${\mathbb R}^3$ with a conformal metric} \label{conformal}
 Suppose now $M$ is ${\mathbb R}^3$ endowed with the metric
\[ds^2=f^2(x_3)(dx_1^2+dx_2^2+dx_3^2).\]
%Let $\sigma =f^3 d_1 \wedge dx_2\wedge dx_3$ be the volume form on $M$.

 It turns out that such choices of $M$ and $ds^2$ provide examples of 3-spaces of arbitrary constant curvature ~\cite{rimgeom}. Indeed, let $K$ be a non-negative real number. If $f(x_3)=1/(K x_3^2+1/4)$, a direct calculations reveals that $M$ has curvature $K$. For $f(x_3)=1/(1+\sqrt{K}x_3)$, we have that $\tilde{M}=\{(x_1,x_2,x_3) \in M / x_3>0 \}$, the open half-space, has curvature $-K$. 
 
 Stokes equations~(\ref{stokes1}) and~(\ref{stokes2}) are
\begin{eqnarray}
\frac{\partial p}{\partial x_1}&=&\eta \frac{2 g'f'}{f^4}, \label{conff1}\\
\frac{\partial p}{\partial x_3}&=&\eta \frac{g''}{f^3}. \label{conff2}
\end{eqnarray}
 
  If we apply the condition~(\ref{maineq}) for existence of functions $g,$ $p$ satisfying the equations above, we obtain
\begin{equation*}
\left(-2\frac{f'}{f^4}\right)'g'=\frac{g'''}{f^3},
\end{equation*}
which is a separable equation.
The differential equation above has a non-trivial solution if and only if $f^3\left(\frac{f'}{f^4}\right)'$ is constant, i.e, the conformal factor must satisfy a differential equation of the form
\begin{eqnarray*}
 f^3\left(\frac{f'}{f^4}\right)'=\tilde{k}_1 & \Rightarrow -\frac{2}{3}f^3(\frac{1}{f^3})''=k_1,
\end{eqnarray*}
By setting $y=\frac{1}{f^3}$, we obtain the equation $y''=-\frac{3}{2}k_1 y$, whose solutions are well-known for any values of the constant $k_1$. For each solution $y=y(x_3)$ there corresponds an $f=y^{-1/3}$ and a $g$ which is a solution of
\begin{equation} \label{conf1}
g'''+k_1 g'=0.
\end{equation} 

 Thus a restriction on the metric needs to be imposed in order to have non-trivial solutions of Stokes equation. Unfortunately, this restriction applies to the choices of conformal factors which give constant curvature. In particular we have that symmetric unidirectional flows are not possible in hyperbolic 3-space.

\section{ The existence of symmetric unidirectional flows}

 Let us address the fundamental question of existence of solutions of equation~(\ref{maineq}).
 
 Recall that the existence of unidirectional flows depends on the existence of a solution $g=g(x_1)$ of equation~(\ref{maineq}). As we mentioned before, equation (\ref{maineq}) is a third order linear ordinary differential equation. It can be put in the form
\[ g'''- A(x_1,x_3)g''-B(x_1,x_3)g'-C(x_1,x_3)g=0. \]
We will show that a differential equation such as this one can only have a solution if the coefficients $A$, $B$, $C$ do not depend on $x_3$. We will need the following 

\begin{lemma} If the linear differential equation
\[ g''-A(x_1,x_3)g'-B(x_1,x_3)g=0 \]
has a solution, then the coefficients $A$ and $B$ do not depend on $x_3$.
\end{lemma}
\dem A solution of the equation in the statement satisfies the linear system
\begin{eqnarray*}
Ag'+Bg&=&g''\\
 A_3g'+B_3g&=&0,
\end{eqnarray*}
where the subscript indicates partial derivative with respect to $x_3$. Thus the $2 \times 2$ determinant 
\begin{equation*}
\left|\begin{array}{cc}
A&B\\
A_3&B_3
\end{array}
\right|
\end{equation*}
is zero. But from $A_3g'+B_3g=0$, we have that $A_3$ is equal to a function $\mu$ of $x_1$ times $B_3$. Hence, we must have $A=\mu B$. So
\[g''=B(\mu g'+g). \]
Therefore $B=B(x_1)$ and $A=A(x_1)$, as we wanted to prove. \vege

\begin{lemma} \label{neccond}
If the linear differential equation
\[ g'''- A(x_1,x_3)g''-B(x_1,x_3)g'-C(x_1,x_3)g=0 \]
has a solution, then the coefficients $A$, $B$ and $C$ do not depend on $x_3$.
\end{lemma}
\dem Consider the linear system
\begin{eqnarray}
Ag''+Bg'+Cg&=&g''', \label{eqa}\\
 \label{eqb} A_3g''+B_3g'+C_3g&=&0 .
\end{eqnarray}
From the previous lemma, equation (\ref{eqb}) implies that there exist functions $\mu$ and $\nu$ of $x_1$ such that $B_3=\mu A_3$ and $C_3=\nu A_3$. Integrating these equations with respect to $x_3$, we obtain
\[
 B= \mu A+\gamma(x_1), \quad C= \nu A+\varpi(x_1).
\]
Substituting into equation (\ref{eqa}), we obtain
\[ A(g''+ \mu g'+\nu g)+(\gamma\ g'+\varpi g)=g'''.\]
Therefore, $A$ is a function of $x_1$ only, and the same must hold for $B$ and $C$. \vege
\fimp

 Lemma \ref{neccond} provides necessary conditions for the existence of symmetrical unidirectional flows on a 3-manifold $M$ with a metric $ds^2$ whose coefficients are $E_1$, $E_2$ and $E_3$. For instance, the coefficient of $g''$ in equation (\ref{maineq}) is
\[ A(x_1,x_3)= -\frac{\partial}{\partial x_1}ln \left(\frac{E_3}{E_1^3 E_2}\right). \]
Lemma \ref{neccond} says that the right-hand side is a function of $x_1$. Hence, a quick calculation shows that $\frac{E_3}{E_1^3 E_2}$ must be separable. We have proved the

\begin{prop} \label{gexis}
 A necessary condition for the existence of a solution of the differential equation (\ref{maineq}) is that $\frac{E_3}{E_1^3 E_2}$ is separable.
\end{prop}
 
 Propositions \ref{stokessep} and \ref{gexis} imply the main result of this section.
 
\begin{thm} \label{necsuf}
 Suppose $E_1=E_1(x_1)$. If Stokes equation has a solution then $\frac{E_3}{E_2}$ is separable. Conversely, if $\frac{E_3}{E_2}$ is separable, then Stokes equation in the metric with coefficients $E_1$, $E_2$ and $E_3$ is separable and hence it has a solution.
\end{thm} 
 
  Theorem \ref{necsuf} implies the impossibility of symmetric unidirectional flows in conical or toroidal geometries.\\
  
\noindent{\bf{\emph{Example 4:}}} Let $M$ be an open region of ${\mathbb R}^3$ endowed with the Euclidean metric in conical coordinates $(\eta,\phi,\rho)$
\[ ds^2=d\eta^2+(\eta \cos \alpha+\rho \sin \alpha)^2 d\phi^2+ d\rho^2,\]
where $0<\alpha<\frac{\pi}{2}$ is a constant.
Since $E_1=1$ and 
\[ \frac{E_3}{E_2}=\frac{1}{\eta \cos \alpha+\rho \sin \alpha} \]
is not separable, we have that equation (\ref{maineq}) has no solution.
\fimp

\noindent{\bf{\emph{Example 5:}}} Let $M$ be an open region of ${\mathbb R}^3$ endowed with the Euclidean metric in toroidal coordinates $(r,\phi,\theta)$
\[ ds^2=dr^2+(a+r\cos \theta)^2d\phi^2+r^2 d\theta^2,\]
where $a>0$ is constant.
Since $E_1=1$ and 
\[ \frac{E_3}{E_2}=\frac{r}{a+r\cos \theta} \]
is not separable, it follows that (\ref{maineq}) has no solution. 
\fimp
 
  If $E_1$ depends on $x_3$, it is possible to have solvable Stokes equations which are not separable, as we saw in subsection 2.2.
  
\begin{rem} 
 Theorem \ref{necsuf} imposes a serious restriction on an argument that has been used to deduce Darcy's law for Hele-Shaw systems in curved geometries. In the next section we discuss a method which provides a perturbed form of Darcy's law which is valid for the example in subsection 2.2, somehow bypassing the obstacle imposed by the non-existence of a symmetric unidirectional flow.
\end{rem}

\section{ Darcy's law}

 We now deduce Darcy's law for a separable and a non-separable example of Hele-Shaw systems.
 
\subsection{Pseudo-Spheres in Minkowski space}

 This example illustrates well the method of deduction of Darcy's law for separable systems. 
 
 Recall that $x_1=r$, $x_2=\phi$ and $x_3=\tau$ are the pseudo-spherical coordinates of $M$ defined in example 3 of subsection 2.1.
 
 Consider the pseudo-spheres $S_a=\{r=a\}$ and $S_{a+b}=\{r=a+b\}$ in $M$. A curve going from $P \in S_a$ to $S_{a+b}$ in the $r$-direction is given by a path $\lambda:[0,1] \rightarrow M$, where 
\begin{eqnarray*}
  r(t)=r(P)+t\,b, & \phi(t)=\phi(P), & \tau(t)=\tau(P).
\end{eqnarray*} 
We  average the function $V_\tau (r,\tau)$ along the path $\lambda$. 

In order to solve equation (\ref{mink}), we set both sides equal to a constant $C$. If we impose the non-slip boundary conditions $g(a)=g(a+b)=0$, then we must have
 \begin{equation} \label{mink1}
  g(r)=-\frac{C}{2\eta} (r-a)(a+b-r). 
\end{equation}
The average of $V_\tau$ along $\vec{\lambda}$ is
\[ \overline{V}_\tau=\frac{\int_{\lambda}{V_\tau}ds}{\int_{\lambda} ds}, \]
where $ds$ is the element of arc-length of $\vec{\lambda}$. We have that
\[ \overline{V}_\tau=\frac{\int_{0}^{1}{V_\tau(\lambda(t))} ib\, dt}{\int_0^1 ib\, dt}=\frac{1}{b \sinh \tau}\int_a^{a+b}\frac{g(r)}{r} dr.\]
Using  expression (\ref{mink1}) for $g$, it follows that
\[ \overline{V}_\tau=-\frac{1}{2 b \eta}\left(\int_a^{a+b}\frac{(r-a)(a+b-r)}{r} dr \right)\frac{\partial p}{\partial \tau}.\]
Therefore, Darcy's law for two pseudo-spheres $S_a$, $S_{a+b}$ in Minkowski's 3-space is
\begin{equation} \label{D1}
 \overline{V}_\tau=-\frac{b^2 \mathcal{F}\left(\frac{b}{a}\right)}{12 \eta}(grad \, p)_\tau ,
\end{equation}
where $\mathcal{F}\left(\frac{b}{a}\right)=F(1,2;4;-b/a)$ is a hypergeometric function. For comparison, see the appendix of~\cite{sphere}.

\subsection{Parallel planes in ${\mathbb R}^3$ with a conformal metric}
 
 We return to the example discussed in subsection (\ref{conformal}).

 Let us consider the planes $S_a=\{x_1=a\}$ and $S_{a+b}=\{x_1=a+b\}$. We define a curve $\lambda:[0,1] \rightarrow {\mathbb R}^3$ by
\begin{eqnarray*}
  x_1(t)=x_1(P)+t\,b, & x_2(t)=x_2(P), & x_3(t)=x_3(P).
\end{eqnarray*} 
where $P \in S_a$. In order to simplify our calculations, we will assume that $a=0$ and $b>0$.

 The average of a function $h:{\mathbb R}^3 \rightarrow {\mathbb R}$ along $\lambda$ is
\[ \overline{h}=\frac{\int_{\lambda} h ds}{\int_{\lambda} ds}=\frac{\int_{0}^{1}h(\lambda(t)) f(x_3(P)) dt}{\int_0^1 f(x_3(P)) b dt}= \frac{1}{b}\int_a^{a+b} h(x_1,x_2(P),x_3(P)) dx_1.\]
We will regard $\,\bar{}\,$ as an averaging operator, which has the property of being linear with respect to functions of $x_3$. Besides, we also have
\begin{equation} \label{parbarr}
\overline{\frac{\partial h}{\partial x_3}}=\frac{\partial \overline{h}}{\partial x_3}.
\end{equation}

 If we apply  $\,\bar{}\,$ to both sides of equations (\ref{g-eq}), (\ref{conff1}) and (\ref{conff2}), we obtain
\begin{eqnarray}
\overline{V}_3&=&\frac{\overline{g}}{f^2}, \label{avevel}\\
\overline{\frac{\partial p}{\partial x_1}}&=& \left(\frac{2\eta f'}{f^4}\right)\overline{g'}, \label{norgrad} \\
\overline{\frac{\partial p}{\partial x_3}}&=& \left(\frac{\eta}{f^3}\right)\overline{g''}. \label{avegrad}
\end{eqnarray}
If we impose the no-slip boundary conditions $g(0)=g(b)=0$, then $\overline{g'}=0$ and hence $\overline{\frac{\partial p}{\partial x_1}}=0$. So, from now on, we will drop equation (\ref{norgrad}). 

 As we have seen in subsection \ref{conformal}, the function $g(x_1)$ must satisfy
\begin{eqnarray}
 g'''+k_1 g'=0 , \label{conff3}
\end{eqnarray}
where $k_1=-\frac{2}{3}f^3\left(\frac{1}{f^3}\right)''$. Equation (\ref{conff3}) is equivalent to 
\[  g''+k_1 g= k_2 \]
for some constant $k_2$. The solution of this equation satisfying the no-slip boundary conditions is
\begin{equation} \label{mainform}
g(x_1)=\frac{k_2}{k_1}\left[1+\left(\frac{\sinh \alpha(x_1-b)-\sinh \alpha x_1}{\sinh \alpha b}\right)\right],
\end{equation}
where $k_1=-\alpha^2$. The averages of $g$ and $g''$ are thus
\begin{eqnarray*}
\overline{g}=\frac{k_2}{k_1}\left[1+2\left(\frac{1-\cosh \alpha b}{\alpha b\sinh \alpha b}\right)\right],& \overline{g''}=\frac{-2 k_2}{\alpha b}\left[\left(\frac{1-\cosh \alpha b}{ \sinh \alpha b}\right)\right].
\end{eqnarray*}
Using these formulas, we obtain 
\[ \overline{V}_3 =\frac{f}{\eta}\left(\frac{\overline{g}}{\overline{g''}}\right)\left(\frac{\partial \overline{p}}{\partial x_3}\right)=\frac{bf }{2 \eta \alpha}\left[\frac{ \sinh \alpha b}{1-\cosh \alpha b}+\frac{2}{\alpha b} \right]\left(\frac{\partial \overline{p}}{\partial x_3}\right).\]
Therefore Darcy's law is 
\[\overline{V}_3 =\frac{bf^2 }{2 \eta \alpha}\left[\frac{ \sinh \alpha b}{1-\cosh \alpha b}+\frac{2}{\alpha b} \right](grad \, \overline{p})_3.\]

 If we substitute into the formula above the power series expressions for the hyperbolic sine and hyperbolic cosine, we get after some manipulation
\begin{equation} \label{fini}
 \overline{V}_3 =-\frac{b^2f^2 }{ \eta }\left[\frac{\left(\frac{1}{3!}-\frac{2}{4!}\right)+\left(\frac{1}{5!}-\frac{2}{6!}\right)\alpha^2b^2+\ldots}{1+2\frac{\alpha^2b^2}{4!}+2\frac{\alpha^4b^4}{6!}+\ldots} \right](grad \, \overline{p})_3, 
\end{equation}
where $\alpha^2=\frac{2}{3}f^3\left(\frac{1}{f^3}\right)''$.

 The important facts about the formula above are
\begin{itemize}
\item [(I)] It is a generalization of Darcy's law for the flat, Euclidean space. Darcy's law in this particular case is obtained by setting $f=1$ (and $\alpha=0$) in (\ref{fini}).
\item [(II)] It is defined for arbitrary conformal factors $f$, even for the ones for which the corresponding symmetric unidirectional flows do not exist!
\end{itemize}

 In the case of a hyperbolic 3-space of curvature $-K$, we have
\[ \alpha^2=4K f^2=\frac{4K}{(1+\sqrt{K}x_3)^2}. \]
Since $x_3>0$, this expression is less than or equal to $4K$, and thus a near-zero choice of curvature will make $\alpha^2$ uniformly small.

\section{Summary and concluding remarks}

In this article we have considered the generalization of fluid flow to non-Euclidean spaces by obtaining Stokes equation for symmetric unidirectional flows in a smooth orientable manifold of dimension 3. We have also found the conditions under which Stokes equation is separable. As examples we recovered Stokes equation in ${\mathbb R}^3$ with the Euclidean metric both in cylindrical and spherical coordinates. This was also done for ${\mathbb R}^3$ with Minkowski metric. We studied then the case of a flow in ${\mathbb R}^3$ with a conformal metric and found that a restriction on the conformal factor is needed in order to have non-trivial solutions of Stokes equation. This restriction rules out spaces of constant curvature such as the hyperbolic 3-space. The existence of symmetric unidirectional flows was addressed and a condition on the manifold metric established for Stokes equation to have solutions. This conditions rules out symmetric unidirectional flows in conical and toroidal geometries. In the conical case, the problem seems to come from the curvature singularity at the cone vertex. In reference \cite{cone} this problem was avoided by cutting out the vertex in order to provide the inlet for the flow. Darcy's law was finally obtained for the cases of two pseudo-spheres in Minkowiski space and for two parallel planes in ${\mathbb R}^3$ with conformal metric. The latter case considers even the case where symmetric unidirectional flows are not possible.  A series expansion of Darcy's law  for small values of the parameter of separation of the two parallel planes recovered Darcy's law in Euclidean space in the unit conformal factor limit. Since Darcy's law is the starting point for the study of key nonlinear aspects of the Saffman-Taylor problem like finger competition and finger tip-splitting, we hope our work will motivate further investigations of such important interfacial features in a variety of curved Hele-Shaw geometries.

\ack
We thank CNPq, PRONEX and CAPES (PROCAD) for financial support. Jos\'e A. Miranda thanks CNPq for a postdoctoral 
scholarship PDE proc. number 200045/2005-9.


\section*{References}
\begin{thebibliography}{10}
 \bibitem{Rev1} Thompson D W 1994 {\it On Growth and Form} (Cambridge: Cambridge University Press)
 \bibitem{Rev2} Stevens P S 1974 {\it Patterns in Nature} (Boston: Little Brown)
 \bibitem{Rev3} Meinhartd H 1982 {\it Models of Bilogical Pattern Formation} (New York: Academic Press)
 \bibitem{Rev4} Koch A J and Meinhardt H 1994 {\it Rev. Mod. Phys.} {\bf 66} 1481
 \bibitem{Rev5} Hyde S, Anderson S, Larsson K, Blum Z, Landh T, Lidin S, and Ninham B W 1997 {\it The Language of Shape. The Role of Curvature in Condensed Matter: Physics, Chemistry and Biology} (Amsterdam: Elsevier)
 \bibitem{saff} Saffman P G and  Taylor G I 1958 {\it Proc. R. Soc. London, Ser. A} {\bf 245} 312
 \bibitem{Ben} Bensimon D, Kadanoff L P, Liang S, Shraiman B I, and Tang C 1996 {\it Rev. Mod. Phys.} {\bf 58} 977
 \bibitem{Hom} Homsy G 1997 {\it Ann. Rev. Fluid Mech.} {\bf 19} 271
 \bibitem{McC} McCloud K V and Maher J V 1995 {\it Phys. Rep.} {\bf 260} 139
 \bibitem{flat} Miranda J A and Widom M 1998 {\it Physica D} {\bf 120} 315
 \bibitem{sphere} Parisio F, Moraes F, Miranda J A and Widom M 2001 {\it Phys. Rev. E} {\bf 63} 036307 
 \bibitem{sphereg} Miranda J A, Parisio F, Moraes F  and  Widom M 2001  {\it Phys. Rev. E} {\bf 63} 63111 
 \bibitem{cyl} Miranda J A 2002 {\it Phys. Rev. E} {\bf 65}  63031 
 \bibitem{cone} Miranda J A and Moraes F 2003  {\it J. Phys. A: Math. and Gen.} {\bf 36}  863; Miranda J A 2002 {\it Phys. Rev. E} {\bf 65}  3101 
 \bibitem{rimgeom} do Carmo M 1992 {\it Riemmanian Geometry} (Boston: Birkh\"auser)
 \bibitem{darling} Darling R W R 1994 {\it Differential Forms and Connections} (Cambridge: Cambridge University Press)
\end{thebibliography}
\end{document}